\documentclass[aps,prl,twocolumn,showpacs]{revtex4}

\usepackage{epsfig}

\usepackage{bbm}

\newcommand{\lbfig}[1]{\refstepcounter{fig} \label{#1} }
\newcounter{fig}

\newcommand{\nc}{\newcommand}
\nc{\be}{\begin{equation}}
\nc{\ee}{\end{equation}}
\nc{\bea}{\begin{eqnarray}}
\nc{\eea}{\end{eqnarray}}
\nc{\bi}[1]{\bibitem{#1}}
\nc{\lsim}{\mbox{\raisebox{-.6ex}{~$\stackrel{<}{\sim}$~}}}
\nc{\gsim}{\mbox{\raisebox{-.6ex}{~$\stackrel{>}{\sim}$~}}}

\begin{document}

\leftline{\hspace{4.4in} CERN-TH/2003-070, HD-THEP-03-15}

\vskip 0.1in

\title{Coherent Baryogenesis}

\author{Bj\"orn Garbrecht$^*$, Tomislav Prokopec$^{*\scriptsize\spadesuit}$
                       and 
                   Michael G. Schmidt
       }
\email[]{B.Garbrecht@thphys.uni-heidelberg.de}
\email[]{T.Prokopec@thphys.uni-heidelberg.de}
\email[]{M.G.Schmidt@thphys.uni-heidelberg.de}

\affiliation{Institut f\"ur Theoretische Physik, Heidelberg University,
             Philosophenweg 16, D-69120 Heidelberg, Germany}

\affiliation{$^\spadesuit$ {\tt Mar 01 - Aug 31 2003:}
             Theory Division, CERN, CH-1211 Geneva 23, Switzerland}


\begin{abstract}
We propose a new baryogenesis scenario based on coherent production
and mixing of different fermionic species. 
The mechanism is operative during phase transitions, at which the
fermions acquire masses {\it via} Yukawa couplings to
scalar fields. Baryon production is efficient 
when the mass matrix is nonadiabatically varying, nonsymmetric
and when it violates $CP$ and $B\!-\!L$ directly, or some other 
charges that are eventually converted to $B\!-\!L$. 
We first consider a toy model, which involves
two mixing fermionic species, and then a hybrid inflationary scenario embedded
in a supersymmetric Pati-Salam GUT. We show that, quite
generically, a baryon excess in accordance with observation can result.
\end{abstract}

\pacs{05.60.Gg, 12.60-i, 98.80.Cq}

\maketitle

%
%

\vskip 0in
{\it 1. Introduction.}
Baryogenesis scenarios beyond the standard model often invoke 
the out-of-equilbrium decay of heavy
particles~\cite{Yoshimura:1978ex,Fukugita:1986hr}.
Well known notable exceptions are the mechanisms proposed by
Affleck and Dine~\cite{Affleck:1984fy} and by
Cohen and Kaplan~\cite{Cohen:1987vi}, both being operative in 
the presence of scalar condensates.

In this letter, we present a novel baryogenesis scenario involving
nonadiabatic evolution of classical scalar fields during phase transitions,
at which the baryon-lepton symmetry of the gauge group
of a grand-unified theory (GUT) is broken. Yukawa couplings
give rise to time dependent mass terms for matter fields and thereby 
the resonant production of particles~\cite{Traschen:1990sw}.
Furthermore, the mass matrices are temporarily nondiagonal,
inducing thus flavor oscillations between species with baryonic
and leptonic charge. At the end of the phase transition,
when the mass matrix becomes diagonal, the charges get frozen-in.
Since this mechanism relies on the interplay of coherent particle
production and $(B\!-\!L)$-violating flavor oscillations
we call it \emph{coherent baryogenesis}. We emphasise the conceptual
simplicity of coherent baryogenesis, since it involves 
the tree-level dynamics of quantum fields only.

\vskip 0.1in

{\it 2. Formalism.}
For our calculations we adapt the kinetic approach
({\it cf.} Refs.~\cite
{ZhuangHeinz:1995,Kainulainen:2001cn,GarbrechtProkopecSchmidt:2002}), 
which makes use of the dynamics of Wigner functions, which are closely
related to classical phase space distributions. 
 
To compute them, we extend
the formalism of Ref.~\cite{Kainulainen:2001cn,GarbrechtProkopecSchmidt:2002}.
We define the Wigner function,
\noindent\vskip -0.15in
\begin{eqnarray}
   iS^<(k,x)_{ab}
 = -\int d^4r {\rm e}^{ik\cdot r}
   \langle 0|\bar{\psi}_{b}(x-r/2)\psi_{a}(x+r/2)|0\rangle
,
\nonumber
\end{eqnarray}
\vskip -0.05in\noindent
where $a,b=1,...,N$ are species indices and
$(i\gamma^0S^{<})^\dagger = i\gamma^0S^{<}$ is hermitian.
Our derivation will go through for a 2-point
function with general density matrix, but in view of our applications in
inflation we prefer to write it with respect to the vacuum from the outset.
Introducing an $N\times N$ matrix $M$
and its hermitian and antihermitian parts, respectively,
\noindent\vskip -0.15in
\begin{eqnarray}
M_H=\frac{1}{2}\left(M+M^{\dagger}\right),\qquad
M_A=\frac{1}{2i}\left(M-M^{\dagger}\right)
,
\nonumber
\end{eqnarray}
\vskip -0.05in\noindent
we find that $iS^{<}$ obeys the Wigner space Dirac equation
\begin{equation}
\Bigl( {k}\!\!\!/\; + \frac{i}{2}\gamma^0 \partial_{t}
    - (M_{H}+i \gamma^5 M_{A})
       e^{-\frac{i}{2}\stackrel{\!\!\leftarrow}{\partial_{t}}\partial_{k_0}}
\Bigr)_{\!\!ac} iS_{cb}^{<} = 0
\label{DiracEq}
.
\end{equation}

The mass matrix $M$ emerges generically from Yukawa couplings to scalar field
condensates, ${\cal L}_{\rm Yu} = -y\phi\bar\psi_R\psi_L + {\rm h.c.}$, 
which can induce $CP$-violation in the fermionic
equations~\cite{JoyceProkopecTurok:1994,ClineJoyceKainulainen:2000,
Carena:2000id,Kainulainen:2001cn}. It is noteable, that in coherent baryogenesis
the condensate does not have to carry a charge, while this is necessary
for the Affleck-Dine mechanism~\cite{Affleck:1984fy}, therefore being
conceptually different. This can be seen explicitly from the two examples
in this letter, where the scalar condensates $\phi$ are always real,
and therefore
$q_\phi=i\langle\phi^{\dagger}
\stackrel{\!\!\leftrightarrow}{\partial_{t}}\phi\rangle=0$.

However, the time dependence of $M$ plays here another important r\^ole.
For $N=1$, the Dirac equation consists of two
first order differential equations 
due to the two degrees of freedom of the fermionic field. 
Thus, also ${d}M/{dt}$ can contribute $CP$-violation, and in general, 
both sources from $M$ and ${d}M/{dt}$ cannot be simultaneously removed 
by local phase
reparametrizations of the fermionic fields,
\emph{cf.} Ref.~\cite{GarbrechtProkopecSchmidt:2002} for an example.
When $N>1$, even higher derivatives of $M$ are involved, allowing,
in principle, multiple sources of $CP$-violation. We stress that 
this is a very different situation from the Standard Model quark mixing
Cabibbo-Kobayashi-Maskawa matrix, or lepton mixing Maki-Nakagawa-Sakata matrix,
where at least three generations of quarks or leptons are required for one
CP-violating phase. 

We now make use of the fact that the helicity operator
$
\hat{h} = \hat{\mathbf{k}}\cdot\gamma^{0}
          \mbox{\boldmath{{$\gamma$}}}\gamma^{5}
$,
commutes with the Dirac operator in (\ref{DiracEq}) and decompose the Wigner
function

\begin{equation}
-i\gamma_{0}S^{<}_{h}
  = \frac{1}{4}\bigl(\mathbbm{1}+h\hat{\mathbf{k}}\cdot
                     \mbox{\boldmath{{$\sigma$}}}
               \bigr)\otimes\rho^{\mu}g_{\mu h}
,
\label{S<:helicity-diagonal}
\end{equation}
\vskip -0.05in\noindent
where we have omitted the species indices,  
$\hat{\mathbf{k}} = \mathbf{k}/|\mathbf{k}|$
and $\sigma^{\mu}$, $\rho^{\mu}$ ($\mu=0,1,2,3$) are the Pauli matrices.
We multiply~(\ref{DiracEq}) by $\rho^{\mu}$, take the trace and integrate
the hermitian part over $k_0$.
Introducing the 0th momenta of $g_{\mu h}$,
 $f_{\mu h}=\int({dk_0}/{2\pi})g_{\mu h}$,
we obtain the system of equations
\begin{eqnarray}
\dot{f}_{0h} + i\left[M_H,f_{1h}\right] +i\left[M_A,f_{2h}\right] &=& 0
\label{f0eq}
\\
\dot{f}_{1h} + 2h|\mathbf{k}|f_{2h} + i\left[M_H,f_{0h}\right] -\left\{M_A,f_{3h}\right\}  &=& 0
\label{f1eq}
\nonumber\\
\dot{f}_{2h} - 2h|\mathbf{k}|f_{1h}  + \left\{M_H,f_{3h}\right\} +i\left[M_A,f_{0h}\right] &=& 0
\label{f2eq}
\nonumber\\
\dot{f}_{3h} - \left\{M_H,f_{2h}\right\} +\left\{M_A,f_{1h}\right\}  &=& 0
\nonumber
.
\end{eqnarray}
The $f_{\mu h}(x,\mathbf{k})$ can be interpreted as follows:
$f_{0h}$ is the charge density, $f_{3h}$ 
is the axial charge density, and $f_{1h}$ and $f_{2h}$ correspond
to the scalar and pseudoscalar density, respectively.
Note that the commutators in~(\ref{f0eq}), which mix particle flavors, 
are essential for the production of the charges $f_{0h}$, and
thus for our scenario. 

For an originally diagonal mass matrix and adiabatic conditions,
the initial Wigner functions describing
zero particles are (\emph{cf.} Ref.~\cite{GarbrechtProkopecSchmidt:2002}):
\begin{eqnarray}
 f^{ab}_{0h} &=& |L_{h}^{ab}|^2+|R_{h}^{ab}|^{2},
\quad 
 f^{ab}_{1h} = -2\Re(L_{h}^{ab}R_{h}^{*ab}),
\nonumber\\
 f^{ab}_{3h} &=& |R_{h}^{ab}|^2-|L_{h}^{ab}|^2,
\quad
 f^{ab}_{2h} = 2\Im(L_{h}^{*ab}R_{h}^{ab})
,
\label{f3asLR}
\end{eqnarray}
with
\vskip -0.25in
\begin{eqnarray}
L_h^{ab}=\delta_{ab}\sqrt{\frac{\omega_{a}+hk}{2\omega_{a}}},\quad
R_h^{ab}=\delta_{ab}\frac{M_{aa}^*}{\sqrt{2\omega_{a}(\omega_{a}+hk)}}
,
\nonumber
\end{eqnarray}
where $\omega_{a}=\sqrt{\mathbf{k}^2+|M_{aa}|^2}$.
For a nondiagonal, but hermitian, $M$, one obtains initial conditions by an
appropriate unitary transformation. If additionally $M_A\not=0$, a biunitary
transformation is necessary for diagonalisation.

Since $f_{0h}$ is the zeroth component
of the vector current, the charge of  the species $a$ carried
by the mode with momentum $k$ and helicity $h$
is simply $q_{ah}(k)\!=\!f_{0h}^{aa}$.
From eqs.~(\ref{f0eq}) one can derive that, in order to generate a
nonvanishing charge 
$q_{ah}(k)$, {\it i.e.} $\dot{f}_{0h}^{aa}\not=0$,
$M$ should not be symmetric.
Note also, that the Lagrangean
\noindent\vskip -0.2in
\begin{eqnarray}
{\cal L}=
\bar{\psi}_{a}\partial\!\!\!/\, \psi_{a} - \bar{\psi}_{b} (M_H + i \gamma^5 M_A)_{ba} \psi_{a}
\nonumber
\end{eqnarray}
\vskip -0.05in\noindent
is $U(1)$ symmetric, and thus $\sum_a q_{ah}(k)$ is conserved.
\vskip 0.1in

{\it 3. Toy model.} 
We now consider a two species model, where fundamental
$SU(2)$ fermions couple to an adjoint scalar triplet $\Phi^i$ and to a 
singlet $\Phi^0$, such that the mass matrix is given by
$M=\Phi^0\mathbbm{1}+\Phi^i\sigma^i$.
While we fix the fields associated with the diagonal
generator $\sigma^3$ as $\Phi^0\!=\!\mu$ and $\Phi^3\!=\!\mu/2$, 
with $\mu$ being a mass scale, we let $\Phi^1$ and $\Phi^2$
move freely in a harmonic potential,
starting form arbitrary initial conditions. Conjugating the fermionic
sector of~(\ref{f0eq}) under $CP$
while leaving the scalar condensate invariant, we find that,
in spite of $M$ being hermitian, $CP$ is broken for the fermions, because 
$M\not=M^*$.
Damping is introduced through a phenomenological decay rate $\Gamma$ 
and through the Hubble expansion 
in a matter dominated universe, \emph{e.g.} the
scale factor is $a=a_m\eta^2$, where $\eta$ denotes the conformal time, and
$'\equiv d/d\eta=ad/dt$.
The equation of motion for a scalar $\Phi(\eta)$ is given by
\begin{equation}
\Phi''+2\frac{a'}{a}\Phi'+a^2\frac{dV}{d\Phi}+a\Gamma\Phi'=0
.
\label{scalareq}
\end{equation}
Writing $\omega_\Phi^2\!\!=\!\!d^2V/d\Phi^2$ and setting $\Gamma\!=\!0$, 
the solutions are
\begin{eqnarray}
\Phi(\eta)=\frac{c_1}{\eta^3} \cos\Big(\frac{1}{3}a_m \omega_\Phi\eta^3\Big)
          + \frac{c_2}{\eta^3} \sin\Big(\frac{1}{3}a_m \omega_\Phi\eta^3\Big)
\label{matter_sol}
.
\end{eqnarray}
For $\Phi^1$, we employ $c_1=\mu$, $c_2=0$ and $\omega_\Phi=\mu$,
for $\Phi^2$, $c_1=0$, $c_2=\mu$ and $\omega_\Phi=1.5\mu$, 
and we set $a_m=\mu^2$.

We approximate the effect of damping
by multiplying the solutions~(\ref{matter_sol}) by
$A\!=\!\exp\left(-\frac{1}{6}a_m\Gamma\eta^3\right)$,
where $\Gamma=0.1\mu\ll\omega_\Phi$.
The equations of motions for the Wigner functions in conformal space-time
are then simply  obtained by replacing $M$ by $aM$ in~(\ref{f0eq})
~\cite{Giudice:1999fb},~\cite{GarbrechtProkopecSchmidt:2002}.
We illustrate the motion of the mixing
fermionic mass terms in conformal time in figure~\ref{figure 1}.
\begin{figure}[htbp]
\begin{center}
\epsfig{file=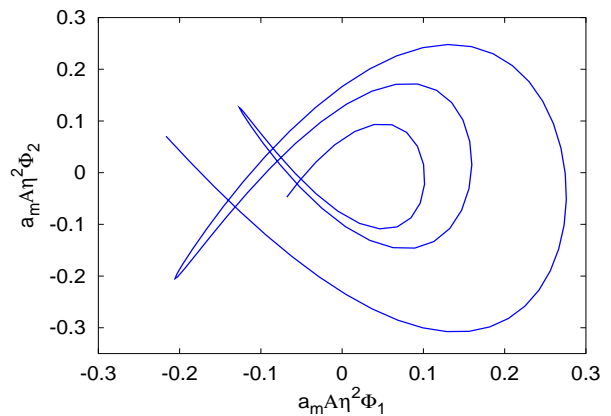, height=2.2in,width=3.3in}
\end{center}
\vskip -0.20in
\caption[fig1]{
\small
Parametric plot of the motion of $a(\eta)A\Phi_1$ and $a(\eta)A\Phi_2$ for
$\eta\in [2.3\mu^{-1},4\mu^{-1}]$.
}
\lbfig{figure 1}
\vskip -0.1in
\end{figure}

Requiring that there are no particles at $\eta\!=\!2.1\mu^{-1}$, 
we choose initial conditions in accordance with~(\ref{f3asLR}) 
and solve (\ref{f0eq}) numerically, to
find fermion number production as displayed in figure~\ref{figure 2} 
as a function of the conformal momentum $k$.
\begin{figure}[htbp]
\begin{center}
\epsfig{file=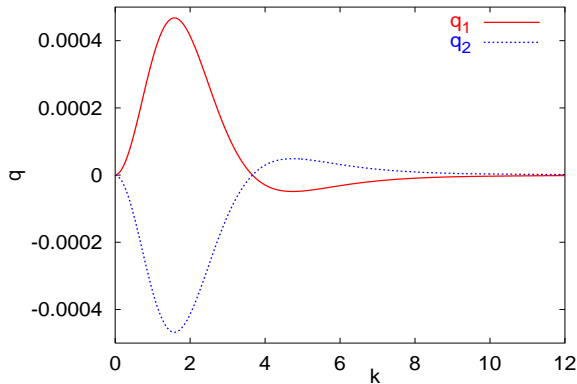, height=2.2in,width=3.3in}
\end{center}
\vskip -0.40in
\caption[fig1]{
\small
The fermion numbers $q_{1+}(k)$ and $q_{2+}(k)$ 
in the toy model.
}
\lbfig{figure 2}
\vskip -0.15in
\end{figure}
We sum over the helicities, $q_a=q_{a+}+q_{a-}$, and
note, that in the present case $q_{a+}=q_{a-}$, for $M_A=0$.
Integration over $k$, $q_{a}=a^{-3}\int dk k^2q_a(k)/2\pi^2 $, 
gives the charge densities
$q_{1}\!=\!-q_{2}=2.4\!\times\!10^{-5}(\mu/a)^3$.
If $q_1$ was charged under $B$, our toy model would lead to 
successful baryogenesis.

\vskip 0.1in

{\it 4. Hybrid inflation in a 
supersymmetric Pati-Salam model.}
We shall now discuss the implementation of
coherent baryogenesis in a more realistic model. 
In order to generate a baryon asymmetry which survives 
the sphaleron washout, we require the presence of ($B\!-\!L$)-violation. 
This is the case in several GUTs, \emph{e.g.} $E(6)$,
a subgroup of which is $SO(10)$, and in the Pati-Salam group,
$G_{PS}=SU(4)_C\times SU(2)_L\times SU(2)_R$,
which appears as an intermediate stage of breaking of $SO(10)$ 
down to the Standard Model. 
For simplicity, here we study an extension of the hybrid inflationary
scenario embedded in a supersymmetric Pati-Salam model, which is considered
in~\cite{Jeannerot:2000sv} and
does not suffer from the monopole problem.
The relevant terms of the superpotential are
\noindent\vskip -0.2in
\begin{eqnarray}
W&\supset&\kappa S \left(\bar{H}^c H^c-\mu^2\right)-
\beta S\left(\frac{\bar{H}^c H^c}{M_S}\right)^2
\nonumber\\
&+&\zeta G H^c H^c+\xi G\bar{H}^c \bar{H}^c
\label{superpotential}
.
\end{eqnarray}
\vskip -0.05in\noindent

Under $G_{PS}$, the fields transform as
$H^c\!=\!(\mathbf{\bar{4}},\mathbf{1},\mathbf{2})$,
$\bar{H}^c\!=\!(\mathbf{4},\mathbf{1},\mathbf{2})$,
$S\!=\!(\mathbf{1},\mathbf{1},\mathbf{1})$,
$G\!=\!(\mathbf{6},\mathbf{1},\mathbf{1})$~\cite{Antoniadis:1988cm}.
We adopt the notation
\noindent\vskip -0.2in
\begin{eqnarray}
H^c=\left(\begin{array}{cccc}
u_{H1}^c & u_{H2}^c & u_{H3}^c & \nu_H^c\\
d_{H1}^c & d_{H2}^c & d_{H3}^c & e_H^c
\end{array}\right)
\nonumber
,
\end{eqnarray}
\vskip -0.05in\noindent
and likewise for $\bar{H}^c$. With $SU(4)_C$ broken to the
Standard Model, $G\!=\!D+\bar{D}$,
with $D\!=\!\mathbf{3}$ and $\bar{D}\!=\!\mathbf{\bar{3}}$ of $SU(3)_C$. 
Vanishing of the D-terms requires $H^c\!=\!\bar{H}^{c*}$,
which we assume throughout
this discussion.
During inflation, the neutrino-like
component $\nu_H^c$ of $H^c$ has the vacuum expectation value
\noindent\vskip -0.2in
\begin{equation}
\left\langle |\nu_H^c| \right\rangle
=M_S \left[\kappa/(2\beta)\right]^{\frac{1}{2}}
\label{infvac},
\end{equation}
\vskip -0.05in\noindent
which evolves during the waterfall regime to the supersymmetric vacuum,
where $S=0$ and
\vskip -0.2in
\begin{eqnarray}
\left\langle |\nu_H^c| \right\rangle=
\biggl(\frac{\kappa M_S^2-
 \left[\kappa M_S^2(\kappa M_S^2-4\beta \mu^2)
 \right]^{\frac{1}{2}}}{2\beta}
\biggr)^{\frac{1}{2}}
\label{supersymmetric vacuum}
.
\end{eqnarray}
All other components of $H^c$ as well as the fields
$G$ vanish at all times. Thus, $|H^c|^2=\bar{H}^c H^c=|\nu_H^c|^2$, and
eq.~(\ref{superpotential}) implies the following scalar potential:
\begin{eqnarray}
V\!\!\!\!&=&\!\!\!2\left|S\nu_H^{c*}\!\left(\!\!\kappa
\!-\!2\beta \frac{|\nu_H^c|^2}{M_S^2}\!\right)\!\right|^2\!\!
+\left|\kappa\left(|\nu_H^c|^2\!-\mu^2\right)\!-\!\beta\frac{|\nu_H^c|^4}{M_S^2}
\right|^2\nonumber
.
\end{eqnarray}

Numerically, we simulate the evolution of the scalar fields from the end of
inflation until they settle to the supersymmetric 
vacuum.
The choice of our parameters is listed in table~\ref{table 1},
for a discussion of their implications for cosmic observations, see
Ref.~\cite{Senoguz:2003hc}.
\begin{table}
\begin{tabular}{|rrl|rrl|rrl|}
\hline
$\ \kappa$ &$=$& $0.007\    $ & $\ \beta   $ &$=$& $1\ $ & $\ \zeta$ &$=$& $0.01\ $
 \\
$\ \Gamma$ &$=$& $0.008\mu\ $ & $\ M_S$ &$=$& $50\mu\ $ & $\ \xi$   &$=$& $0.01\!\times\!\exp\left(i\!\times\! 10^{-3}\right)\ $
 \\
\hline
\end{tabular}
\caption[tab1]{
\small
Parameters used in numerical simulation.
}
\label{table 1}
\vskip -0.15in
\end{table}
We let $\nu_H^c$ point in the real direction and remark,
that not all constants
can be made real by field redefinitions, thereby allowing a 
nontrivial phase between $\zeta$ and $\xi$.
Note, that the substitution $\zeta\longleftrightarrow \xi^*$
results in the opposite sign of $B\!-\!L$ and $B$ and hence,
when $\zeta=\xi^*$, there is no generation of $B\!-\!L$.
Furthermore, that $M_S$ is rather small is not
an essential feature of the model, 
but was introduced to avoid large initial Higgs field amplitudes, 
which are unplesant to deal with numerically.

\begin{figure}[htbp]
\vskip -0.15in
\begin{center}
\epsfig{file=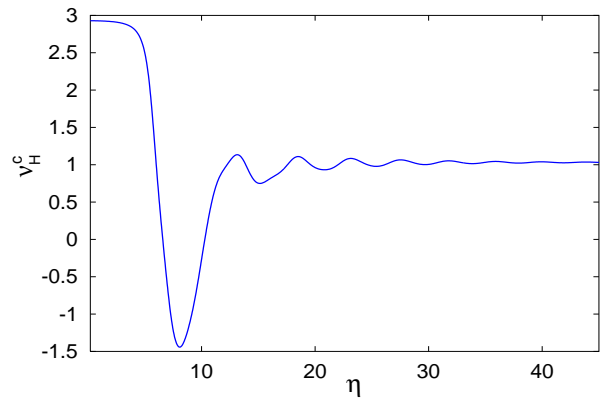, height=2.2in,width=3.3in}
\end{center}
\vskip -0.35in
\caption[fig1]{
\small
Evolution of the field $\nu_H^c$.
}
\lbfig{figure 3}
\vskip -0.10in
\end{figure}

Eventually, $\nu_H^c$ condenses to the value given 
by~(\ref{supersymmetric vacuum})~({\it cf.} figure~\ref{figure 3}), 
and the field $S$ settles to zero.
In this situation, the flavor oscillations between the Dirac fermions
%
%

%
\noindent\vskip -0.2in
\begin{eqnarray}
\chi_{1j}=\left(\begin{array}{c}
-\psi_{d_{\bar{H}j}^c} \\ \bar{\psi}_{\bar{D}_j}
\end{array}\right)
\ \textnormal{and} 
\ \chi_{2j}=\left(\begin{array}{c}
-\psi_{D_j} \\ \bar{\psi}_{d^c_{Hj}}
\end{array}\right)
,
\label{chi_12}
\end{eqnarray}
\vskip -0.05in\noindent
are scotched, since their mass term,

\noindent\vskip -0.2in
\begin{eqnarray}
\left(\begin{array}{cc}\bar{\chi}_{1j} & \bar{\chi}_{2j}\end{array}\right)
\Bigg[\!\!\!
&\left(\begin{array}{cc}
\Re\left[{\langle \nu_H^c \rangle \xi}\right] & \frac{1}{2}m_d\\
\frac{1}{2}m_d &\Re\left[{\langle\nu_H^{c} \rangle\zeta}\right]
\end{array}\right)&\nonumber
\\
+i\gamma^5
&\left(\begin{array}{cc}
-\Im\left[{\langle \nu_H^c \rangle \xi}\right] & -\frac{i}{2}m_d\\
\frac{i}{2}m_d &-\Im\left[{\langle \nu_H^{c} \rangle \zeta}\right]
\end{array}\right)&
\!\!\!\Bigg]
\left(\begin{array}{c}\chi_{1j} \\ \chi_{2j}\end{array}\right)
\nonumber
,
\end{eqnarray}
\vskip -0.05in\noindent
where $m_d\!=\!\langle S \rangle
       \left(\kappa/2  - \beta \langle |\nu_H^c| \rangle ^2/M_S^2\right)$,
becomes diagonal.

Starting with initially zero fermions, we compute the
charges $q_{1j}(k)$ and $q_{2j}(k)$ associated with 
$\chi_{1j}$ and $\chi_{2j}$, respectively, at the time when
the phase transition is completed, by using~(\ref{f0eq}), 
as shown in figure~\ref{figure 4}. Integration then yields
$q_1\!=\!-q_2\!=\!1.0\!\times\!10^{-5}(\mu/a)^3$, where we have
already summed over the three colors, and where
in our computation $a\approx 120$ during the phase transition.

\begin{figure}[htbp]
\begin{center}
\epsfig{file=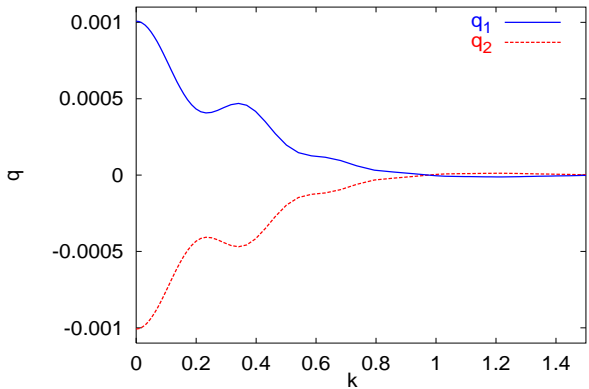, height=2.2in,width=3.3in}
\end{center}
\vskip -0.35in
\caption[fig1]{
\small
The produced charges 
of the Dirac fermions $\chi_{1j}$, $\chi_{2j}$,
summed over both helicities.
}
\lbfig{figure 4}
\vskip -0.1in
\end{figure}

In the presence of the $(B\!-\!L)$-violating
condensate of $\bar\nu_H^c$, the right handed neutrino $\nu^c$
acquires a Majorana
mass from the additional, nonrenormalizable contribution
$
\gamma F^c\bar{H}^cF^c\bar{H}^c/M_S
$
to the superpotential~(\ref{superpotential}), where $F^c\bar{H}^c$
form a gauge singlet, and
$F^c=(\mathbf{\bar{4}},\mathbf{1},\mathbf{2})$ are the superfields containing
the right handed quarks and leptons,
\noindent\vskip -0.2in
\begin{eqnarray}
F^c=\left(\begin{array}{cccc}
u_{1}^c & u_{2}^c & u_{3}^c & \nu^c\\
d_{1}^c & d_{2}^c & d_{3}^c & e^c
\end{array}\right)
\nonumber
.
\end{eqnarray}
\vskip -0.1in\noindent
This also gives rise to the Lagrangean terms
\noindent\vskip -0.2in
\begin{eqnarray}
\gamma(\langle\bar{\nu}_H^c\rangle/M_S)[\psi_{\nu^c}\phi_{d_j^c}\psi_{d_{\bar{H}j}^c}
+\psi_{d_j^c}\phi_{\nu^c}\psi_{d_{\bar{H}j}^c}]+c.c.
\nonumber
,
\end{eqnarray}
\vskip -0.1in\noindent
allowing the decay of the $d_{\bar{H}j}^c$-component
of $\chi_{1j}$ in~(\ref{chi_12}) to
$d^{c*}\!+\nu^{c*}$,
where one of the latter particles is fermionic, 
the other scalar.
The Majorana neutrinos $\nu^c$ are their own antiparticles; 
neglecting the small effects induced by possible mixing angles and a
$CP$-violating phase in the Majorana mass matrix familiar from
leptogenesis scenarios~\cite{Fukugita:1986hr,Buchmuller:1996pa,Giudice:1999fb,Garcia-Bellido-Morales:2001},
their decay leaves behind no net charge. Note that in turn, our
mechanism does not require any lower constraint on these angles 
and the phase as in leptogenesis.

The coupling
$
\gamma_2 F^cH^cF^cH^c/M_S
$
allows the decay of the $H^c$-fields through
the reaction from the $d_{H}^{c*}$-component of
$\chi_{2j}$ in~(\ref{chi_12}) to
$d^{c}\!+u^{c}$.
The charges hence get transformed to
$\left(B\!-\!L\right)\!=\!\frac{1}{3} q_1\!-\!\frac{2}{3}q_2\!$.
This number is promoted by sphaleron processes to
$B=({10}/{31})(B\!-\!L)\!$~\cite{Harvey:1990qw},
where we assumed two complex Higgs doublets.

Hence, the final value of $B-L$ arises here
due to the transformation of other charges in decay processes.
However, we point out, that models are conceivable where 
coherent particle production directly leads to 
standard model particles, the $\left(B\!-\!L\right)$-charge
of which is conserved in the  subsequent history of the universe.


To obtain a lower bound on the baryon-to-photon ratio, we 
need an upper bound for the entropy,
which can be obtained by assuming a complete and instantaneous conversion
of the vacuum energy 
$\varrho\!=\![\kappa^2M_S^2/(4\beta)-\kappa\mu^2]^2$
into a thermal bath of highly relativistic particles, with the energy density
$\varrho\!=\!\pi^2 g^* T^4/30$, where 
$g^*\!=\! 221.5$ is the number of degrees of freedom in the MSSM.
The entropy per unit volume is then $s=2 \pi^2 g^* T^3/45$, and
we thus estimate the generated baryon-to-photon ratio to be
$B/n_\gamma \simeq 9.4\!\times\!10^{-10}$.
Also, a contribution from boson production may arise, which we 
do not consider here. Note that our result was obtained by choosing
a small CP-violating phase ({\it cf.} table~\ref{table 1}), 
indicating that there is an ample phase space of couplings,
which leads to baryon production consistent with the observed
$B/n_\gamma\!=\!6.1\pm 0.3\!\times\!10^{-10}$~\cite{Bennett:2003bz}. 
In conclusion, we have demonstrated that coherent baryogenesis is a 
viable, efficient and natural candidate for the creation of the
baryon asymmetry of the universe.


%
%

\end{document}